\documentclass[a4paper,fleqn,usenatbib,useAMS]{mnras}

\usepackage{newtxtext,newtxmath}

\usepackage[T1]{fontenc}
\usepackage{ae,aecompl}

\usepackage[dvipdfmx]{graphicx}	
\usepackage{amsmath}	
\usepackage{amssymb}	
\usepackage{bm}
\usepackage{color}

\title[Cosmic Ray Modified Shock Polarimetry]{Diagnosing Cosmic Ray Modified Shocks with H $\alpha$ Polarimetry}

\author[J. Shimoda \& J. M. Laming]{
Jiro Shimoda$^{1,2}$\thanks{E-mail: j-shimoda@astr.tohoku.ac.jp (JS)}
and J. Martin Laming $^3$\thanks{E-mail: laming@nrl.navy.mil (JML)}
\\
$^{1}$Frontier Research Institute for Interdisciplinary Sciences, Tohoku University, Sendai 980-8578, Japan\\
$^{2}$Astronomical Institute, Tohoku University, Sendai 980-8578, Japan\\
$^{3}$Space Science Division Code 7684, Naval Research Laboratory, Washington DC 20375, USA\\
}

\date{Accepted XXX. Received YYY; in original form ZZZ}

\pubyear{2018}

\begin{document}
\label{firstpage}
\pagerange{\pageref{firstpage}--\pageref{lastpage}}
\maketitle

\begin{abstract}
A novel diagnostic of cosmic-ray modified shocks by polarimetry of
H~$\alpha$ emissions is suggested. In a cosmic-ray modified shock, the
pressure of cosmic rays is sufficiently high compared to the upstream ram
pressure to force the background plasma to decelerate (measured in the
shock rest frame). Simultaneously, a fraction of the hydrogen atoms
co-existing in the upstream plasma collide with the decelerated protons and
undergo charge-exchange reactions. As a result, hydrogen atoms with the
same bulk velocity of the decelerated protons are generated. We show that
when the shock is observed from edge-on, the H~$\alpha$ radiated by these
upstream hydrogen atoms is linearly polarized with a sizable degree of a
few per cent as a result of resonant scattering of Ly~$\beta$. The
polarization direction depends strongly on the velocity modification; the
direction is parallel to the shock surface for the case of no modification,
while the direction is parallel to the shock velocity for the case of a
modified shock.
\end{abstract}

\begin{keywords}
acceleration of particles
-- atomic processes
-- radiative transfer
-- shock waves
-- cosmic rays
-- ISM: supernova remnants.
\end{keywords}



\section{Introduction}
Supernova remnant (SNR) shock waves are believed to be the origin of Galactic
cosmic-rays (CRs). If this is the case, they should produce a sufficient
amount of CRs to explain the CR energy density measured in the vicinity of
the Earth (e.g. $\simeq1~{\rm eV~cm^{-3}}$ for CR protons) and should
accelerate CR-nuclei up to at least $10^{15.5}~{\rm eV}$ (the so-called knee
energy). The former can be estimated by measurements of the energy loss rate
of shocks due to CR acceleration~\citep[e.g.][and references
therein]{shimoda18,hovey18}. The latter is one of the most actively discussed
topics using combinations of wide band observations such as $\gamma$-ray,
X-ray and radio wave~\citep[e.g.][for {\it Tycho}'s SNR and references
therein]{archambault17,laming15}. Although such investigations are widely
reported on, these two requirements are still open issues.
\par
From a theoretical point view, the acceleration efficiency of CRs and the
maximum energy of CRs realized in SNR shocks are related to each other. The
most generally accepted mechanism for CR acceleration is diffusive shock
acceleration~\citep[DSA, e.g.][]{bell78,blandford78}. In the DSA mechanism,
particles bouncing back and forth between the upstream and downstream regions
are accelerated. This behaviour is quantified by their spatial diffusion
coefficient, which depends on the nature of magnetic disturbances~\citep[see,
e.g.][for review]{blandford87}.
In order to accelerate CR nuclei to the knee energy in SNR shocks, an
extremely disturbed magnetic field with a strength of $\sim100~{\rm \umu G}$
is required in the upstream region~\citep[e.g.][]{lagage83a,lagage83b}.
However, the typical strength of the magnetic field in the local interstellar
medium (upstream region) is estimated to be at most
$\sim10~{\rm \umu G}$~\citep[e.g.][]{myers78,beck01}. Motivated by this
issue of magnetic field strength, \citet{bell04} provided a linear analysis
of parallel shocks efficiently accelerating CR protons. He showed that the
upstream magnetic field is amplified by the effects of a back reaction from
the accelerated protons themselves. This amplification is called the Bell
instability whose growth rate is proportional to the square root of the CR
energy density. On the other hand, it is also known that the CRs leaking to
the upstream region can have a comparable pressure compared to the ram
pressure of the upstream background protons. As a result, measured in the
rest frame of shock, the background thermal protons are decelerated by the
pressure gradient due to CRs {\it before} entering the shock
front~\citep[e.g.][]{drury81,drury86}. Such a shock is called a CR modified
shock. Hence, when the CR acceleration efficiency is sufficiently high, in
other words, when the CR pressure or energy density is large enough, the DSA
mechanism is regulated by the CRs themselves via the effects of the back
reaction on the background plasma~\citep[see also][for more
detail]{laming14}.
\par
The CR modified shock scenario discussed above is often employed in SNR
shocks to explain the spectral energy distribution (SED) of non-thermal
emissions~\citep[see, e.g.][]{archambault17}. However, current SED fitting
can not discriminate between the acceleration models. For example, whether
TeV band $\gamma$-rays originate from the CR nuclei or CR electrons is
representative of the problem.
This issue mainly depends on the uncertainties of structure of ambient medium
and uncertainties of the details of plasma physics including the back reaction
rather than the quality of the observational SED data. Indeed, \citet{inoue12} and
\citet{inoue19} showed that the multiphase structure of the interstellar
medium as a consequence of thermal instability results in the same form of
$\gamma$-ray spectrum for the cases of CR nuclei and CR electrons.
Thus, we need other constraints for further investigations of CR acceleration in SNR
shocks.
\par
Observation of the velocity modification of upstream protons, which has
unfortunately never been done, would provide us with a novel insight into the
CR acceleration process because the degree of modification and the length
scale of the modified region would reflect the CR acceleration efficiency and
the diffusion length of CRs respectively~\citep[see, e.g.][for a simplified
case]{berezhko99}. Note that this means that we should know the condition of
the background {\it upstream} plasma rather than the CRs themselves to reveal
the physics of the acceleration. The plasma conditions around the shock can
be investigated through observations of the hydrogen Balmer lines, especially
H~$\alpha$~\citep[e.g.][]{chevalier80,raymond91,smith94}. Indeed, the
H~$\alpha$ emissions from the upstream region have been observed in actual
SNRs~\citep[e.g.][]{lee07,katsuda16,knezevic17}. This upstream H~$\alpha$ can
originate from the absorption of Ly~$\beta$ photons radiated from the
downstream region~\citep[e.g.][]{shimoda19}. The absorption of Ly~$\beta$
results in an excited hydrogen atom in the 3p state followed by the excited
atom emitting H~$\alpha$ due to the transition from 3p to 2s with a
probability of 0.12 (otherwise it re-emits Ly~$\beta$). Note that this is a
scattering of an electromagnetic wave and therefore the H~$\alpha$ should
have a net linear polarization. For the case of a CR modified shock, the
decelerated protons can interact with hydrogen atoms and be changed to
hydrogen atoms by the charge-exchange reaction, H$+$p$\rightarrow$p$+$H,
where H and p denote the hydrogen atom and the proton, respectively~\citep{ohira10}. Since
these newly generated hydrogen atoms have a different mean velocity compared
to pre-existing hydrogen atoms, properties of the Ly~$\beta$-H~$\alpha$
conversion in the upstream region differs from the case of no modification.
In this paper, we show that the velocity modification can be measured by
polarimetry of H~$\alpha$: linear polarization parallel to the shock surface
is observed for the case of no CR, while polarization perpendicular to the
surface is observed for the case of a modified shock.
\par
This paper is organized as follows: in Section~\ref{sec:scattering}, we
provide basic concepts for the linear polarization degree and its direction
for H~$\alpha$ resulting from the Ly~$\beta$ scattering. In
section~\ref{sec:model}, we estimate the polarization of H~$\alpha$ for the
cases of no CR precursor, only an electron heating precursor, and a proton
decelerated precursor accompanied by electron heating based on the
sophisticated atomic population calculations of~\citet{shimoda19}. Finally,
we summarize our results and discuss future prospects.
\section{Basic concepts for Polarization of Scattered Line}
\label{sec:scattering} Here we consider the polarization of H~$\alpha$
resulting from the scattering of Ly~$\beta$ following the Rayleigh scattering
formulas \citep[e.g.][]{chandra60}.\footnote{Since the frequency is changed
by the scattering, it is formally called Raman scattering. But because the
polarization results from the conservation of angular momentum rather than
the conservation of energy, the Rayleigh scattering regime can describe the
polarization in Raman scattering.} We treat the radiation transfer problem in
a steady state slab geometry. The slab corresponds to the SNR shock front.
The shock is plane-parallel to the $x$-$y$ plane and axially symmetric about
the $z$-axis. Thus, the specific intensity $I_{\rm \nu,\mu}$ at a position
$z$ depends only on the direction of propagation making an angle $\theta$ to
the $z$-axis. The label of the direction is $\mu\equiv\cos\theta$ and $\nu$
is the frequency of the photon.
\par
%
\begin{figure*}
\includegraphics[scale=0.4]{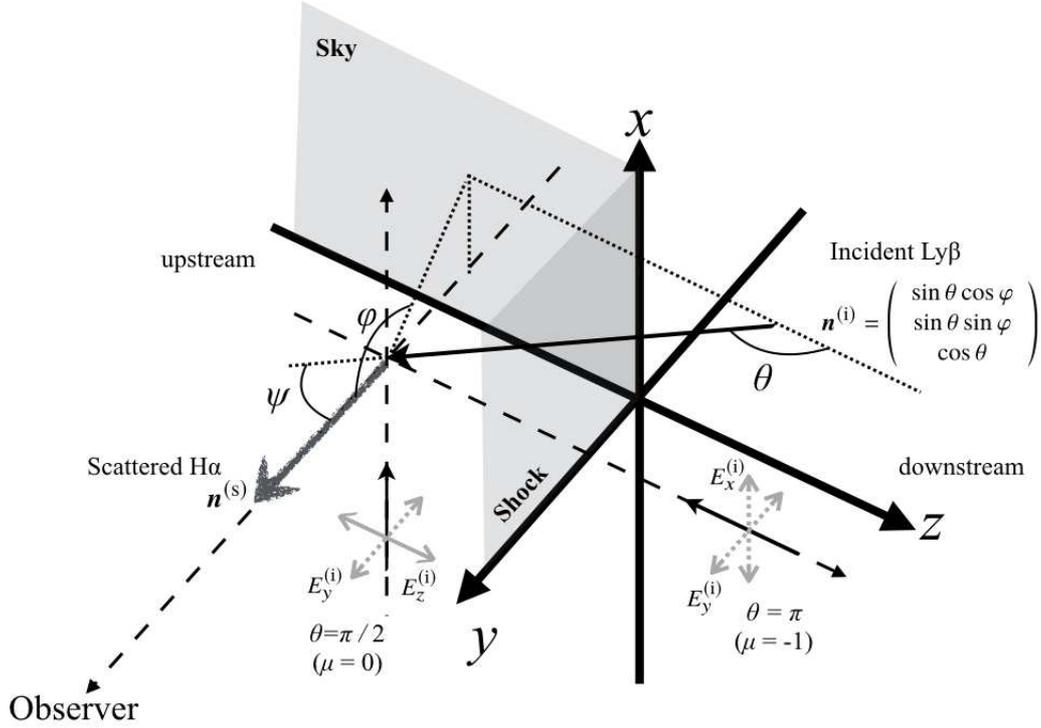}
\caption{Schematic diagram of an SNR shock. The shock is plane and parallel to the $x$-$y$ plane.
We observe the shock from the $y$ direction. The two gray sheets represent the shock surface and the sky.
The thick black arrows indicate the incident Ly~$\beta$ which propagates in the direction of
$\bm{n}^{\rm (i)}=(\sin\theta\cos\varphi,\sin\theta\sin\varphi,\cos\theta)$.
The fuzzy arrow indicates the scattered H~$\alpha$ in the direction of $\bm{n}^{\rm (s)}=(0,1,0)$.
$\bm{n}^{\rm (i)}$ makes an angle $\psi$ with $\bm{n}^{\rm (s)}$. The gray two-way arrows associated with
the incident Ly~$\beta$ in the direction $\theta=\upi/2$ and $\theta=\upi$ indicate the components of electric field.
The dotted two-way arrows indicate the $x$ or $y$ component which may responsible for the excitation of 3p$_{3/2}$ state,
resulting in the negative polarization (along the $x$-axis). The solid two-way arrow indicates the $z$ component which remains
before and after the scattering, giving the positive polarization (along the $z$-axis). The broken and dotted lines are auxiliary lines.}
\label{fig:coordinate}
\end{figure*}
%
Figure~\ref{fig:coordinate} shows a schematic diagram of the shock.
Ly~$\beta$ is incident from the direction of
%
\begin{eqnarray}
\bm{n}^{\rm (i)}=
\left(
\begin{array}{c}
\sin\theta\cos\varphi \\
\sin\theta\sin\varphi \\
\cos\theta
\end{array}
\right),
\end{eqnarray}
%
and it is scattered as H~$\alpha$ in the direction of
%
\begin{eqnarray}
\bm{n}^{\rm (s)}=
\left(
\begin{array}{c}
0 \\
1 \\
0
\end{array}
\right).
\end{eqnarray}
%
The scattering angle is defined as
%
\begin{eqnarray}
\cos\psi=\bm{n}^{\rm (i)}\cdot\bm{n}^{\rm (s)}=\sin\theta\sin\varphi.
\end{eqnarray}
%
In this paper, we fix our line of sight along the $y$-axis so the $z$-$x$
plane corresponds to the sky. Thus, the linear polarization of scattered
H~$\alpha$ is characterized by its $z$ and $x$ components of electric field,
$E_z^{\rm (s)}$ and $E_x^{\rm (s)}$. In addition, we suppose the optically
thin limit for H~$\alpha$. This can be valid for an SNR shock propagating
into a medium with a number density of $<30~{\rm cm^{-3}}$~\citep{shimoda19}.
The polarization is described by the intensity components
[erg~cm$^{-2}$~s$^{-1}$~Hz$^{-1}$~str$^{-1}$]:
%
\begin{eqnarray}
I_{\nu,z}^{\rm (s)}
&\equiv&
a\left< E_{z}^{\rm (s)*} E_{z}^{\rm (s)} \right>,
\label{eq:Iz}
\end{eqnarray}
%
and,
%
\begin{eqnarray}
I_{\nu,x}^{\rm (s)}
&\equiv&
a\left< E_{x}^{\rm (s)*} E_{x}^{\rm (s)} \right>,
\label{eq:Ix}
\end{eqnarray}
%
where the asterisk $^*$ indicates the complex conjugate, $\left<...\right>$
means a long time average and the factor $a$ is a dimensional positive
constant whose actual value is not important in this paper. Then, the Stokes
parameters are defined as
%
\begin{eqnarray}
I_{\nu}^{\rm (s)}
&\equiv&
I_{\nu,z}^{\rm (s)}+I_{\nu,x}^{\rm (s)},
\label{eq:def Stokes I}
\end{eqnarray}
%
%
\begin{eqnarray}
Q_{\nu}^{\rm (s)}
&\equiv&
I_{\nu,z}^{\rm (s)}-I_{\nu,x}^{\rm (s)}
\label{eq:def Stokes Q}
\end{eqnarray}
%
and,
%
\begin{eqnarray}
U_{\nu}^{\rm (s)}
&\equiv&
 a\left< E_{z}^{(s)*} E_{x}^{(s)} \right>
-a\left< E_{x}^{(s)*} E_{z}^{(s)} \right>.
\label{eq:def Stokes U}
\end{eqnarray}
%
Here we omit the label $\mu$ because our line of sight is fixed along the
$y$-axis. A positive $Q_{\nu}^{\rm (s)}$ indicates a linear polarization
along the $z$-axis (perpendicular to the shock surface), while a negative
$Q_{\nu}^{\rm (s)}$ indicates polarization along the $x$-axis (parallel to
the shock surface). For the incident Ly~$\beta$, we use similar notations.
Note that since we consider linear polarization only, the Stokes $V$ is
omitted.
\par
In Rayleigh scattering, the polarization of scattered light is given with
respect to the scattering plane that is defined by $\bm{n}^{\rm (i)}$ and
$\bm{n}^{\rm (s)}$. We denote the component parallel to this plane as
$\parallel$ and the component perpendicular to it as $\perp$. Then, we obtain
the scattered intensities as~\citep{chandra60}
%
\begin{eqnarray}
\left(
\begin{array}{c}
I_{\nu,\parallel}^{\rm (s)} \\
I_{\nu,\perp}^{\rm (s)} \\
\tilde{U}_\nu^{\rm (s)}
\end{array}
\right)
= N_{\rm R} \int {\cal R}
\left(
\begin{array}{c}
I_{\nu,\mu,\parallel}^{\rm (i)} \\
I_{\nu,\mu,\perp}^{\rm (i)} \\
\tilde{U}_{\nu,\mu}^{\rm (i)}
\end{array}
\right)
{\rm d}\Omega^{\rm (i)},
\label{eq:pol 1}
\end{eqnarray}
%
where ${\rm d}\Omega^{\rm (i)}=\sin\theta{\rm d}\theta{\rm d}\varphi$ is the
solid angle element for the incident Ly~$\beta$. The superscript (i)
indicates the incident Ly~$\beta$. The Stokes $U$ of the scattered H~$\alpha$
(incident Ly~$\beta$) are $\tilde{U}_\nu^{\rm (s)}$ ($\tilde{U}_\nu^{\rm
(i)}$) and are defined by the components of electric field $\parallel$ and
$\perp$. The normalization factor we discuss later in Section~\ref{sec:model}
is $N_{\rm R}$. The phase matrix ${\cal R}$ gives the conservation of angular
momentum between the photons and bound electron before and after the
scattering (i.e. the polarization), which depends on the properties of atomic
transitions in general. In our case, the Ly~$\beta$-H~$\alpha$ conversion,
the relevant atomic transitions are from 1s$_{1/2}$ to 3p$_{1/2}$ and to
3p$_{3/2}$ of hydrogen atoms. Note that the subscript $j$ (1/2 and 3/2)
denotes the excited bound electron's angular momentum. According to
\citet{hamilton47} and \citet{chandra60}, the matrices for the cases of
3p$_{1/2}$ ($\Delta j=1/2-1/2=0$) and 3p$_{3/2}$ ($\Delta j=3/2-1/2=1$) are
given by
%
\begin{eqnarray}
{\cal R}=
\left(
\begin{array}{ccc}
1 & 1 & 0 \\
1 & 1 & 0 \\
0 & 0 & 0
\end{array}
\right)
\label{eq:R 1/2}
\end{eqnarray}
%
and
%
\begin{eqnarray}
{\cal R}=\frac{3}{4}
\left(
\begin{array}{ccc}
\cos^2\psi & 0 & 0 \\
    0        & 1 & 0 \\
    0        & 0 & \cos\psi
\end{array}
\right)
+\frac{1}{4}
\left(
\begin{array}{ccc}
1 & 1 & 0 \\
1 & 1 & 0 \\
0 & 0 & 0
\end{array}
\right),
\label{eq:R 3/2}
\end{eqnarray}
%
respectively. We have to transform the polarization with respect to the
scattering plane given by Eq.~\eqref{eq:pol 1} to one with respect to the
$z$-$x$ plane.
Let $\tilde{\theta}$ be an angle between the $z$-axis and
the axis along the parallel component of electric field:
%
\begin{eqnarray}
\tan\tilde\theta
=\frac{\sin\tilde\theta}{\cos\tilde\theta}
=\frac{\sin\theta\cos\varphi}{\cos\theta}.
\end{eqnarray}
%
Note that since the scattering plane includes $\bm{n}^{\rm (s)}=(0,1,0)$, the
parallel and perpendicular axes are always in the $z$-$x$ plane. The parallel
component of electric field is written as
$\bm{E}_{\parallel}=E_{\parallel}\left(\cos\tilde\theta,\sin\tilde\theta\right)$.
Similarly,
$\bm{E}_{\perp}=E_{\perp}\left(\sin\tilde\theta,\cos\tilde\theta\right)$.
Thus, we obtain the intensities measured along the $z$-axis and $x$-axis,
respectively,
%
\begin{eqnarray}
I_z = I_\parallel\cos^2\tilde\theta+I_\perp\sin^2\tilde\theta, \\
I_x = I_\parallel\sin^2\tilde\theta+I_\perp\cos^2\tilde\theta,
\end{eqnarray}
%
and
%
\begin{eqnarray}
U = \left(I_\parallel-I_\perp\right)\sin2\tilde\theta+\tilde{U}\cos2\tilde\theta,
\end{eqnarray}
%
where we omit the notations $\nu,$ $\mu$, (s) and (i).
\par
Hydrogen atoms in the 3p$_{1/2}$ state yield completely unpolarized
H~$\alpha$: $I_{\nu,z}^{\rm (s)}=I_{\nu,x}^{\rm (s)}$ and  $U_{\nu}^{\rm
(s)}=0$. Note that the polarization of a photon can be characterized by the
specific component of its angular-momentum (i.e. spin) along its direction of
travel, and is responsible for the transverse components of the electric
fields. For example, we define that a completely left-handed (right-handed)
circularly polarized photon propagating along the $z$ axis has a
$z-$component of $+1$ ($-1$). In general, the light is partially polarized
because not all photons are emitted with the same angular momentum
$z$-component. The transitions from 1s$_{1/2}$ to 3p$_{1/2}$ are induced by
the absorption equal numbers of left and right circularly polarized photons
with a net component of zero, and the subsequent transitions from 3p$_{1/2}$
to 2s$_{1/2}$ emit photons with the zero net-component so as to satisfy the
conservation of angular momentum. Thus, the scattered H~$\alpha$ is
completely unpolarized in this case.
\par
In the following, we assume that the incident Ly~$\beta$ is always completely
unpolarized for simplicity \citep[c.f.][]{laming90,shimoda18}. Thus,
%
\begin{eqnarray}
I_{\nu,\mu,\parallel}^{\rm (i)}=I_{\nu,\mu,\perp}^{\rm (i)}=\frac{ I_{\nu,\mu}^{\rm (i)}}{2}
\end{eqnarray}
%
and
%
\begin{eqnarray}
\tilde{U}_{\nu,\mu}^{\rm (i)}=0.
\end{eqnarray}
%
Note that 88 per cent of hydrogen atoms in the 3p state emit Ly~$\beta$ again
(not H~$\alpha$). This re-emitted Ly~$\beta$ can be polarized and will be
immediately re-absorbed by other hydrogen atoms. After repeating a few times
this Ly~$\beta$-Ly~$\beta$ scattering, the 3p state hydrogen atom emits
H~$\alpha$ (the Ly~$\beta$-H~$\alpha$ conversion occurs). During this
trapping of Ly~$\beta$, the net polarization of Ly~$\beta$ may be washed out.
Hence, we follow the assumption that all of the incident Ly~$\beta$ is
completely unpolarized.~\footnote{\citet{yang17} calculated a radiation transfer with
Rayleigh scattering by spherical grains for a simple slab geometry and showed
the polarization degree of scattered light from the slab is at most $\simeq1.4$ per cent.
Thus, in our case, the scattered Ly~$\beta$ may be polarized with a degree of $\sim1$ per
cent in reality. We suppose this polarization is negligible.}
\par
For the case of 3p$_{3/2}$ ($\Delta j=1$),
the polarization of H~$\alpha$ is given by
%
\begin{eqnarray}
I_{\nu}^{\rm (s)}
=N_{\rm R}\int_{-1}^{1} I_{\nu,\mu}^{\rm (i)}
\left(
\frac{17}{16}-\frac{3}{16}\mu^2
\right){\rm d}\mu,
\label{eq:Stokes I}
\end{eqnarray}
%
%
\begin{eqnarray}
Q_{\nu}^{\rm (s)}
=N_{\rm R}\int_{-1}^{1} I_{\nu,\mu}^{\rm (i)}
\left(
\frac{3}{16}-\frac{9}{16}\mu^2
\right){\rm d}\mu,
\label{eq:Stokes Q}
\end{eqnarray}
%
and $U_\nu^{\rm (s)}=0$, where we use the notation $\mu=\cos\theta$ and
include the numerical factor coming from the integration by $\varphi$ in
$N_{\rm R}$. Since the specific component of angular momentum varies, the
scattered H~$\alpha$ can be polarized, unlike the case of 3p$_{1/2}$.
The sign of $Q_\nu^{\rm (s)}$ refers whether the polarization is along the
$z$-axis (positive) or the $x$-axis (negative).
\par
We discuss the polarization properties of scattered H~$\alpha$ for the
transition from 1s$_{1/2}$ to 3p$_{3/2}$. First, when all of the photons are
incident from the direction $|\mu|=1$ (Figure~\ref{fig:coordinate} shows the
case of $\mu=-1$), the Stokes $Q_{\nu}^{\rm (s)}$ becomes negative (polarized
along the $x$-axis). This is the usual polarization property for the
scattering of electromagnetic (transverse) waves. Note that this situation
gives a maximum degree of polarization $\simeq -0.43$. On the other hand, for
the case of $\mu=0$ (Figure~\ref{fig:coordinate} shows the case of incident
along the $x$-axis), we observe a positive polarization with $\simeq0.18$
degree. This is also the usual property of electromagnetic wave scattering.
The polarization degree reflects the nature of angular momentum in the atomic
transition ~\citep[the laws of conservation, commutation and synthesis, see
e.g.][for details]{hamilton47}.~\footnote{ In other words, it results from
the bound electron in the photon interactions having discontinuous
eigenvalues. Contrary to this, in the Thomson scattering for example, the
free electron has the continuous eigenvalues and thus results in the (almost)
complete polarization as described by the classical electromagnetism.}
\par
Inside a uniform, isotropically emitting medium the incident intensity is,
%
\begin{eqnarray}
I_{\nu,\mu}^{\rm (i)}=S_\nu(1-{\rm e}^{-\frac{\tau_{\nu}}{|\mu|}}),
\label{eq:simple intensity}
\end{eqnarray}
%
where $S_\nu$ and $\tau_{\nu}$ are $\mu$--independent source function and
optical depth respectively. The Stokes parameters are
%
\begin{eqnarray}
I_{\nu}^{\rm (s)}
&=&N_{\rm R}\int_{-1}^{1} S_\nu (1-{\rm e}^{-\frac{\tau_{\nu}}{|\mu|}})
\left(
\frac{17}{16}-\frac{3}{16}\mu^2
\right){\rm d}\mu, \nonumber \\
&=&N_{\rm R}S_\nu
\left[
2-\int_{-1}^{1} {\rm e}^{-\frac{\tau_{\nu}}{|\mu|}}\left(
\frac{17}{16}-\frac{3}{16}\mu^2
\right){\rm d}\mu
\right],
\label{eq:Stokes I simple}
\end{eqnarray}
%
and
%
\begin{eqnarray}
Q_{\nu}^{\rm (s)}
&=&N_{\rm R}\int_{-1}^{1} S_\nu(1-{\rm e}^{-\frac{\tau_{\nu}}{|\mu|}})
\left(
\frac{3}{16}-\frac{9}{16}\mu^2
\right){\rm d}\mu, \nonumber \\
&=&-N_{\rm R}S_\nu
\int_{-1}^{1} {\rm e}^{-\frac{\tau_{\nu}}{|\mu|}}\left(
\frac{3}{16}-\frac{9}{16}\mu^2
\right){\rm d}\mu.
\label{eq:Stokes Q simple}
\end{eqnarray}
%
Thus, the polarization degree is
%
\begin{eqnarray}
\frac{Q_\nu^{\rm (s)}}{I_\nu^{\rm (s)}}=
\frac{
-\int_{-1}^{1} {\rm e}^{-\frac{\tau_{\nu}}{|\mu|}}\left(
\frac{3}{16}-\frac{9}{16}\mu^2
\right){\rm d}\mu}
{
2-\int_{-1}^{1} {\rm e}^{-\frac{\tau_{\nu}}{|\mu|}}\left(
\frac{17}{16}-\frac{3}{16}\mu^2
\right){\rm d}\mu}.
\label{eq:degree simple}
\end{eqnarray}
%
It can be expressed as
%
\begin{eqnarray}
\frac{Q_\nu^{\rm (s)}}{I_\nu^{\rm (s)}}&=&
\frac{
  -  3 A(\tau_\nu) + 9 B(\tau_\nu)}{
16- 17 A(\tau_\nu) + 3 B(\tau_\nu)}, \\
A(\tau_\nu)&=&{\rm e}^{-\tau_\nu} - \tau_\nu E_1(\tau_\nu), \\
B(\tau_\nu)&=&\frac{{\rm e}^{-\tau_\nu}}{6}\left(\tau_\nu{}^2-\tau_\nu+2 \right)
        - \frac{\tau_\nu{}^3 E_1(\tau_\nu)}{6},
\end{eqnarray}
%
where
%
\begin{eqnarray}
E_1(\tau_\nu)=\int_{\tau_\nu}^{\infty} \frac{{\rm e}^{-\alpha}}{\alpha} {\rm d}\alpha,
\end{eqnarray}
%
is the exponential integral.
%
\begin{figure}
\includegraphics[scale=0.7]{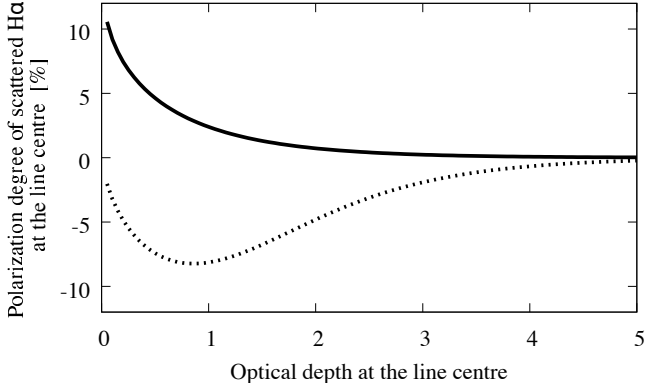}
\caption{The polarization degree of scattered H~$\alpha$ at the line centre.
Note that we consider only the transition from 3p$_{3/2}$ to 2s$_{1/2}$ here.
The solid line is the case of a uniform, isotropically emitting medium
in which the intensity of Ly~$\beta$ is given by Eq.~\eqref{eq:simple intensity}.
The dots represent the case of an anisotropic radiation field modeling the SNR shock.
Here we set $I_{\nu,{\rm B}}=10 S_\nu$ in Eq.~\eqref{eq:mimick}.}
\label{fig:simple}
\end{figure}
%
Figure~\ref{fig:simple} shows the polarization degree at the line centre (see
solid line). Since the intensity of Ly~$\beta$ incident from $\mu\approx0$ is
stronger than from $\mu\approx1$ at small $\tau_\nu$, in other words, since
the incident `photon beam' is elongated along the $x$-axis, the polarization
degree is positive. For $\tau_\nu\gg1$, the radiation field becomes isotropic
(i.e. $I_{\nu,\mu}^{\rm (i)}\approx S_\nu$) and thus the polarization is
zero.
\par
As discussed above, the polarization of the scattered line depends on the
anisotropy of the radiation field. Even if we consider a uniform,
isotropically emitting medium, the anisotropy arises from the effects of
attenuation and the scattered line has net polarization. For SNR shocks,
hydrogen atoms in the upstream region are illuminated by Ly~$\beta$ radiated
from the downstream region. Thus, the radiation field is quite anisotropic at
the shock \citep[e.g. Figure 7 of][]{shimoda19}. This situation may be
modeled by
%
\begin{eqnarray}
I_{\nu,\mu}^{\rm (i)}=I_{\nu,{\rm B}}{\rm e}^{-\frac{\tau_\nu}{|\mu|}}
+S_\nu(1-{\rm e}^{-\frac{\tau_{\nu}}{|\mu|}}),
\end{eqnarray}
%
where $I_{\nu,{\rm B}}$ is the $\mu$-independent intensity of Ly~$\beta$ from
the downstream region, and $\tau_{\nu}$ is now the optical depth outside the
shock. Then, the polarization degree becomes
%
\begin{eqnarray}
\frac{Q_\nu^{\rm (s)}}{I_\nu^{\rm (s)}}&=&
\frac{
  \left( I_{\nu,{\rm B}}-S_{\nu} \right)
  \left(  3 A(\tau_\nu) - 9 B(\tau_\nu)\right)}{
16 S_\nu + \left( I_{\nu,{\rm B}}-S_{\nu} \right)
           \left( 17 A(\tau_\nu) - 3 B(\tau_\nu)\right)}.
\label{eq:mimick}
\end{eqnarray}
%
Figure~\ref{fig:simple} shows the case of $I_{\nu,{\rm B}}=10 S_\nu$ (see
dots), giving `negative' polarization. Note that if the value of $S_\nu$ is
absolutely equal to zero, the polarization degree becomes $-0.43$ at
$\tau_\nu\rightarrow\infty$. In other words, if $S_\nu$ has a very small but
finite value, the degree becomes zero at which $I_{\nu,{\rm B}}{\rm
e}^{-\frac{\tau_\nu}{\mu}}\ll S_\nu$. Hence, in actual SNR shocks, H~$\alpha$
can have a sizable, `negative' polarization degree in the upstream region.
\par
For the CR modified shock, the upstream hydrogen atoms can be decelerated
before entering the shock front as a consequence of charge-exchange reactions
with decelerated protons. This `deceleration' is measured in the shock rest
frame. In the observer rest frame (far upstream rest frame), the particles
are accelerated towards the far upstream region where the CR back reactions
cease. Thus, the atoms in the far upstream region `see' blue-shifted
Ly~$\beta$ radiated by the decelerated atoms. As a result, the optical depth
depends strongly on the direction cosine $\mu$. Because the Ly~$\beta$
incident from $\mu\approx0$ are seen as not shifted, the scattering of the
$\mu\approx0$ photons becomes dominant, giving a `positive' polarization.
\par
In actual SNR shocks, unfortunately, the radiation line transfer problem
including the polarization is quite complex even if there is no shock
modification. Hence, using the latest radiative line transfer model of SNR
shocks constructed by \citet{shimoda19} and making several simplifications,
we estimate the polarization of H~$\alpha$ for both cases of
modified/unmodified shock in this paper.

\section{Model}
\label{sec:model}
%
\begin{figure}
\includegraphics[scale=0.35]{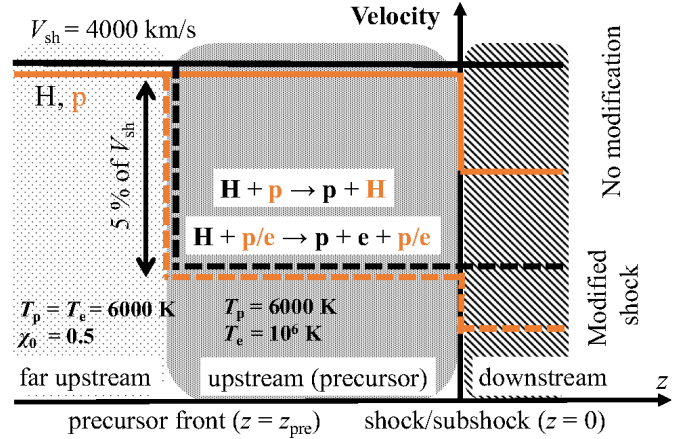}
\caption{Schematic illustration of the shock models considered in this paper.
The vertical axis shows the velocities of hydrogen atoms and protons measured in the shock rest frame.
The horizontal axis is the $z$ axis.
The shock (or sub-shock for the modified shock case) is located at $z=0$.
The solid lines indicate the cases of no velocity modification,
while the broken lines indicate the case of the modified shocks.
The orange lines are the mean velocity of protons and the black lines are the mean velocity of hydrogen atoms, respectively.
The precursor front is located at $z=z_{\rm pre}$ at which the electrons are heated up to $10^6$~K
and the protons are being decelerated to have a mean velocity of $0.95V_{\rm sh}$.
Some of hydrogen atoms having the same bulk velocity as the protons
emerge from charge-exchange reactions.}
\label{fig:models}
\end{figure}
%
We estimate the polarization of H~$\alpha$ from the upstream region of SNR
shocks with fixed shock velocity, $V_{\rm sh}=4000~{\rm km~s^{-1}}$, far
upstream temperature, $T_0=6000$~K, and preshock ionization degree,
$\chi_0=0.5$, for the following three cases: (i) no particles leaking to the
upstream region which is the same situation as in \citet{shimoda19}, (ii) an
electron heating precursor with temperature of $10^6$~K due to CRs  but
no velocity modifications following \citet{laming14}, (iii) in addition
to the electron heating precursor, there are decelerated protons, but with no
proton heating. Figure~\ref{fig:models} summarises these three cases. We
assume that the velocity modification is $5$ per cent of $V_{\rm sh}$ (i.e.
$200~{\rm km~s^{-1}}$) and that the downstream values are given by usual the
Rankine-Hugoniot relations for simplicity. Note that the downstream electron
temperature is fixed at 10 per cent of proton temperature ($T_{\rm
e}\simeq3~{\rm keV}$).
\par
In the radiation line transfer model of \citet{shimoda19}, the shock geometry
is the same as in Figure~\ref{fig:coordinate}. For the radiation transfer
problem, they set the two free escape boundaries for photons upstream and
downstream of the shock. The plasma they consider consists of protons,
electrons and hydrogen atoms. Note that bremsstrahlung radiation, emission
from the SNR ejecta and any other external radiation sources are neglected.
The excitation levels of hydrogen atom are considered up to 4f. Their model
did not consider the existence of a precursor. Therefore, we additionally
solve the ionization structure of hydrogen for the precursor cases.
\par
For case (ii), with only an electron heating precursor, we solve the
ionization structure of the upstream hydrogen in the shock rest frame with
the following equations
%
\begin{eqnarray}
\frac{\upartial n_{\rm H}^0}{\upartial z}=-n_{\rm H}^0\frac{C_{\rm I}}{V_{\rm sh}},
\end{eqnarray}
%
and
%
\begin{eqnarray}
\frac{\upartial n_{\rm p}}{\upartial z}=
n_{\rm H}^0\frac{C_{\rm I}}{V_{\rm sh}},
\end{eqnarray}
%
where $n_{\rm H}^0$ is the number density of hydrogen atoms that have not
experienced the charge-exchange reactions. The number density of protons is
$n_{\rm p}$ and $C_{\rm I}$ is the ionization rate. For the modified shock
case (iii), the ionization structure in the precursor region is given by
%
\begin{eqnarray}
\frac{\upartial n_{\rm H}^0}{\upartial z}=-n_{\rm H}^0\frac{C_{\rm I}+C_{\rm CX}}{V_{\rm sh}},
\label{eq:H0iii}
\end{eqnarray}
%
%
\begin{eqnarray}
\frac{\upartial n_{\rm H}^1}{\upartial z}
=\frac{n_{\rm H}^0 C_{\rm CX} - n_{\rm H}^1 C_{\rm I}}{u_1},
\label{eq:H1iii}
\end{eqnarray}
%
and
%
\begin{eqnarray}
\frac{\upartial n_{\rm p}}{\upartial z}=
\left( n_{\rm H}^0+n_{\rm H}^1 \right)\frac{C_{\rm I}}{u_1}.
\end{eqnarray}
%
where $n_{\rm H}^1$ is the number density of hydrogen atoms that have
experienced charge-exchange reactions with the decelerated protons. The bulk
velocity of the decelerated protons and that of the hydrogen atoms having
undergone charge-exchange reactions is $u_1$. The rate of charge-exchange
reactions for the relative velocity of $200~{\rm km~s^{-1}}$ is $C_{\rm
CX}\simeq4.59\times10^{-8}(n_{\rm p}/{\rm 1~cm^{-3}})~{\rm
s^{-1}}$~\citep{janev93,janev03}. In the following, the total number density
of hydrogen atoms is denoted as $n_{\rm H}=n_{\rm H}^0+n_{\rm H}^1$. Note
that $n_{\rm H}^1=0$ for the cases (i) and (ii) because there are no
decelerated protons. The ionization structure of the downstream region is
given by the same formulations as in \cite{shimoda19}. Note that the velocity
distribution function of all particles are shifted Maxwellians.
%
\begin{figure}
\includegraphics[scale=0.7]{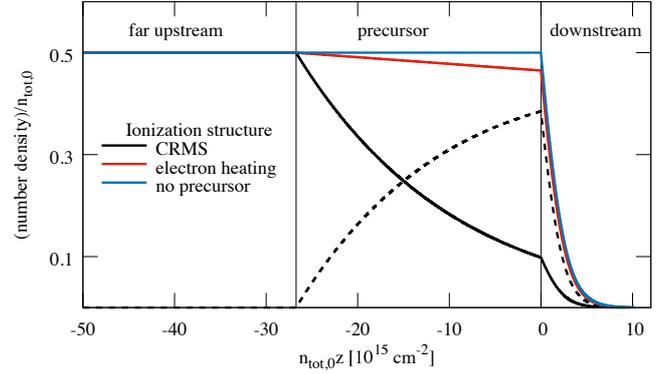}
\caption{The ionization structure of the hydrogen atoms.
The blue, red and black lines indicate the cases of no precursor (i),
only the electron heating precursor (ii), and the electron heating precursor with decelerated protons (iii),
respectively.
The solid lines are the number density of hydrogen atoms that have not experienced a charge-exchange
reaction in the precursor region, $n_{\rm H}^{\rm 0}$.
The broken line is the number density of hydrogen atoms that have experienced a charge-exchange reaction
with the decelerated protons in the precursor region, $n_{\rm H}^{\rm 1}$.
The vertical thin lines at $n_{\rm tot,0} z\simeq-27\times10^{15}~{\rm cm^{-2}}$
and $n_{\rm tot,0} z\simeq0~{\rm cm^{-2}}$ indicate the locations of the precursor front
and the shock front, respectively.}
\label{fig:ionization}
\end{figure}
%
Figure~\ref{fig:ionization} shows the ionization structure of hydrogen. The
results are normalized by the total number density $n_{\rm tot,0}=n_{\rm
p}+n_{\rm H}$. Here we set the precursor front at $n_{\rm tot,0}z_{\rm
pre}=0.14 V_{\rm sh}/\tilde{C}_{\rm I}$, where $\tilde{C}_{\rm I}=C_{\rm
I}/n_{\rm p}\simeq2.15\times10^{-9}~{\rm cm^3~s^{-1}}$ is estimated for
$T_{\rm e}=10^6$~K. The hydrogen atoms in the precursor region are modestly
ionized by the heated electrons but most of them experience charge-exchange
reactions with the decelerated protons ($n_{\rm tot,0}z_{\rm pre}\simeq3
n_{\rm p}V_{\rm sh}/C_{\rm CX}$).
\par
In order to constrain the velocity modification of shock via the H~$\alpha$
emission, the length scale of modification $z_{\rm pre}$ has to be
sufficiently larger than the characteristic length of the spatial
distribution of the decelerated hydrogen atoms emerging from the
charge-exchange reaction with the decelerated protons, $V_{\rm sh}/C_{\rm
CX}(\Delta V_{\rm sh})$, where $\Delta V_{\rm sh}\simeq200~{\rm km~s^{-1}}$
is the relative velocity of collision particles. In this paper, we show the
case of $z_{\rm pre}\simeq3 (n_{\rm p}/n_{\rm tot,0})V_{\rm sh}/C_{\rm CX}$.
If the modification comes from CRs accelerated at the shock, the length scale
of modification can be given by their diffusion length $l_{\rm diff}\sim
D(E)/V_{\rm sh}$, where $D$ and $E$ are the CR diffusion coefficient and
energy, respectively. Thus, from the observational condition $l_{\rm
diff}\ga3 (n_{\rm p}/n_{\rm tot,0})V_{\rm sh}/C_{\rm CX}$, we obtain
%
\begin{eqnarray}
D(E)  &\ga&  1.05\times10^{25}~{\rm cm^{-2}~s^{-1}} \\
&\times& \left( \frac{V_{\rm sh}}{4000~{\rm km~s^{-1}}}        \right)^2
         \left( \frac{n_{\rm tot,0}}{1~{\rm cm^{-3}}}          \right)^{-1}
         \left( \frac{\Delta V_{\rm sh}}{200~{\rm km~s^{-1}}}  \right)^{-1},
\nonumber
\end{eqnarray}
%
where we approximate
%
\begin{eqnarray}
\frac{ C_{\rm CX} }{ n_{\rm p} }
\approx4.59\times10^{-8}~{\rm cm^{3}~s^{-1}}
\left( \frac{\Delta V_{\rm sh}}{200~{\rm km~s^{-1}}} \right).
\nonumber
\end{eqnarray}
%
The diffusion coefficient for the interstellar medium, $D_{\rm ISM}(1~{\rm
GeV})\sim10^{28}~{\rm cm^2~s^{-1}}$, satisfies this condition. When magnetic
disturbances are excited around an SNR shock, we can not currently derive the
relevant diffusion coefficient. If we suppose the simplest case, $D(E)\simeq
\xi_{\rm B}r_g(E)c/3$, where $r_g$ is the gyro radius of CR and $\xi_{\rm
B}>1$ is the gyro factor characterized by the energy spectrum of magnetic
disturbances, we obtain a lower bound for the CR energy $E$ responsible for
the velocity modification of
%
\begin{eqnarray}
E &\ga& 33~{\rm TeV}\frac{1}{\xi_{\rm B}}
\times
\left( \frac{B}{100~{\rm \umu G}}        \right)
\left( \frac{V_{\rm sh}}{4000~{\rm km~s^{-1}}}        \right)^2 \nonumber \\
&\times&
\left( \frac{n_{\rm tot,0}}{1~{\rm cm^{-3}}}          \right)^{-1}
\left( \frac{\Delta V_{\rm sh}}{200~{\rm km~s^{-1}}}  \right)^{-1}.
\end{eqnarray}
%
Note that this estimate of $E$ is based on the coefficient at the Bohm limit,
$D(E)\simeq r_g(E)c/3$, that gives the shortest diffusion length for a given
$E$. For {\it Tycho}'s SNR, it is implied by the two-point correlation
analysis of synchrotron intensity that the energy spectrum of downstream
magnetic disturbances has a Kolmogorov-like
scaling~\citep{shimoda18b}.~\footnote{ \citet{roy09} analyzed the synchrotron
correlation in the SNR~Cas~A but they did not take sufficient care of the
effects of SNR geometry that are the bottleneck in this
type of analysis~\citep[see, e.g.][]{shimoda18b}. Note that to analyze the
synchrotron correlation, \citet{roy09} used interferometric data directly
(i.e. the data are represented in the Fourier space), a different approach to
that of \citet{shimoda18b}. If we removed the geometrical effects from such
an analysis method, we could obtain the energy spectra in the Fourier space.
In any way, it has been suggested that the magnetic energy spectrum can be
measured in principle.  } If this is also true for the upstream disturbances,
the gyro factor $\xi_{\rm B}$ may be greater than unity and the lower bound
of $E$ may be smaller than $33$~TeV. Note that the acceleration efficiency of
CRs can be measured simultaneously by polarimetry of the downstream
H~$\alpha$ emission~\citep{shimoda18}.

\par
We calculate the atomic populations in the same manner as \citet{shimoda19}.
Then, we obtain the emissivity of H~$\alpha$ as
%
\begin{eqnarray}
j_{\nu,\mu}=
\frac{h\nu}{4\upi}
\left(
 n_{\rm H,3s} A_{\rm 3s,2p}\phi_{\nu,\mu}^{\rm 3s,2p}
+n_{\rm H,3p} A_{\rm 3p,2s}\phi_{\nu,\mu}^{\rm 3p,2s}
\nonumber \right. \\ \left.
+n_{\rm H,3d} A_{\rm 3d,2p}\phi_{\nu,\mu}^{\rm 3d,2p}
\right),
\label{eq:emissivity}
\end{eqnarray}
%
where $n_{{\rm H},k}$ denotes the number density of hydrogen atoms in the
excited state $k$. The spontaneous transition rate and line profile function
for the transition from $k$ to $j$ are given by $A_{k,j}$ and
$\phi_{\nu,\mu}^{k,j}$, respectively~\citep[see][for details]{shimoda19}.
Here we omit the term for the two-photon decay. This emissivity gives the
total intensity of H~$\alpha$ including all contributions such as collisional
excitation, radiative excitation and cascades from higher levels (up to 4f in
our model). Almost all radiative excitation comes from the absorption of
Ly~$\beta$, which populates $n_{\rm H,3p}$.
%
\begin{figure}
\includegraphics[scale=0.7]{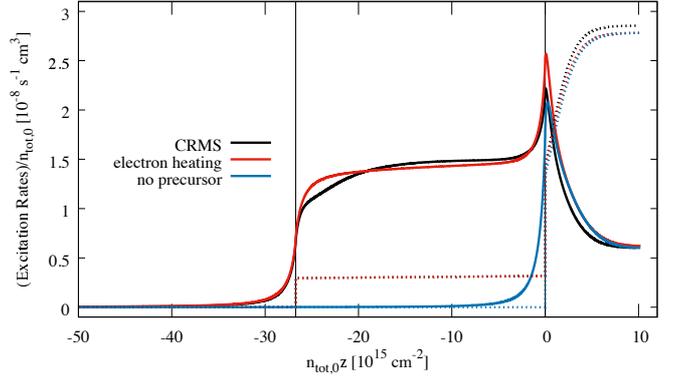}
\caption{
Radiative excitation rate for the transition from 1s to 3p (solid lines)
and total collisional excitation rate for the transitions from 1s to 3s, 3p and 3d (dots).
The blue, red and black lines indicate the cases of no precursor (i),
only an electron heating precursor (ii), and the electron heating precursor with decelerated protons (iii),
respectively.
The vertical thin lines at $n_{\rm tot,0} z\simeq-27\times10^{15}~{\rm cm^{-2}}$
and $n_{\rm tot,0} z\simeq0~{\rm cm^{-2}}$ indicate the locations of the precursor front
and the shock front, respectively.}
\label{fig:crpr}
\end{figure}
%
Figure~\ref{fig:crpr} shows the radiative rate for the transition from 1s to
3p and the total collisional rate of the transitions from 1s to 3s, 3p and
3d. The radiative excitation dominates over the collisional excitation in the
upstream region because the Lyman lines are trapped. Thus, in the upstream
region, most of H~$\alpha$ photons arise from the Ly~$\beta$-H~$\alpha$
conversion. In this paper, for calculating the Stokes $Q$ of H~$\alpha$, we
approximate that all of the 3p-state hydrogen atoms arise from Ly~$\beta$
absorption. Hence, we obtain
%
\begin{eqnarray}
n_{\rm H,3p}=n_{\rm H,3p_{3/2}}+n_{\rm H,3p_{1/2}}
\approx\left(1+\frac{f_{\rm 1s,3p_{1/2}}}{f_{\rm 1s,3p_{3/2}}}\right)n_{\rm H,3p_{3/2}},
\end{eqnarray}
%
where $f_{\rm 1s,3p_{1/2}}=0.026381$ and $f_{\rm 1s,3p_{3/2}}=0.052761$ are
the oscillator strengths for the transitions from 1s to 3p$_{1/2}$ and to
3p$_{3/2}$, respectively~\citep{wiese09}. Since we suppose the optically thin
limit for H~$\alpha$, the Stokes $I$ of the scattered H~$\alpha$ is given by
$I_\nu^{\rm (s)}=j_{\nu,0}^{\rm 3p_{3/2},2s}L$, where $L$ is a path length
along the line of sight and $j_{\nu,\mu}^{k,j}$ is the emissivity for the
transition from $k$ to $j$. Note that our line of sight is fixed along the
$y$-axis ($\mu=0$). Thus, we obtain the normalization factor of the Rayleigh
scattering as
%
\begin{eqnarray}
N_{\rm R}=
\frac{ \frac{h\nu}{4\upi}n_{\rm H,3p_{3/2}}A_{\rm 3p,2s}\phi_{\nu,\mu}^{\rm 3p,2s} L }
     { \int_{-1}^{1}I_{\nu,\mu}^{\rm (i)}
       \left( \frac{17}{16}-\frac{3}{16}\mu^2 \right)  {\rm d}\mu }.
\end{eqnarray}
%
Here we estimate the intensity of Ly~$\beta$ in the scattering as
%
\begin{eqnarray}
I_{\nu,\mu}^{\rm (i)}(z)
=I_{\nu,\mu}^{\rm Ly\beta}(z){\rm exp}\left[{-\frac{k_{\nu,\mu}(z)\Delta z}{|\mu|}}\right],
\end{eqnarray}
%
where $k_{\nu,\mu}$ is the absorption coefficient and $I_{\nu,\mu}^{\rm
Ly\beta}$ is the specific intensity of Ly~$\beta$. They are the outcome of
the atomic population calculations. The spatial resolution of our numerical
calculation is $\Delta z$: $k_{\nu,\mu}\Delta z\simeq0.08$ at the far
upstream region. The polarization degree of H~$\alpha$, finally, is written
as
%
\begin{eqnarray}
\frac{Q_\nu^{\rm (s)}}{I_{\nu,0}}
&=&
\frac{N_{\rm R}\int_{-1}^{1} I_{\nu,\mu}^{\rm (i)} \left( \frac{3}{16}-\frac{9}{16}\mu^2 \right) {\rm d}\mu}
     {j_{\nu,0}L} \nonumber \\
&=&
\frac{ n_{\rm H,3p}A_{\rm 3p,2s}\left( 1 + \frac{ f_{\rm 1s,3p_{1/2}} }{ f_{\rm 1s,3p_{3/2}} } \right)^{-1} }
     { n_{\rm H,3s}A_{\rm 3s,2p}+n_{\rm H,3p}A_{\rm 3p,2s}+n_{\rm H,3d}A_{\rm 3d,2p} } \nonumber \\
&\times&
\frac{ \int_{-1}^{1} I_{\nu,\mu}^{\rm Ly\beta} {\rm e}^{-\frac{k_{\nu,\mu}}{|\mu|}\Delta z }
       \left(\frac{ 3}{16}-\frac{9}{16}\mu^2 \right){\rm d}\mu }
     { \int_{-1}^{1} I_{\nu,\mu}^{\rm Ly\beta} {\rm e}^{-\frac{k_{\nu,\mu}}{|\mu|}\Delta z }
       \left(\frac{17}{16}-\frac{3}{16}\mu^2 \right){\rm d}\mu }.
\end{eqnarray}
%
Note that this polarization degree includes the effects of collisional
excitation, cascades from higher levels and
scattering in the case of $\Delta j=0$ (transitions
from 1s$_{1/2}$ to 3p$_{1/2}$ and subsequently to 2s$_{1/2}$).
In particular, we assume that the precursor collisional excitation
and cascades yield completely unpolarized H~$\alpha$.
\par
%
\begin{figure}
\includegraphics[scale=0.7]{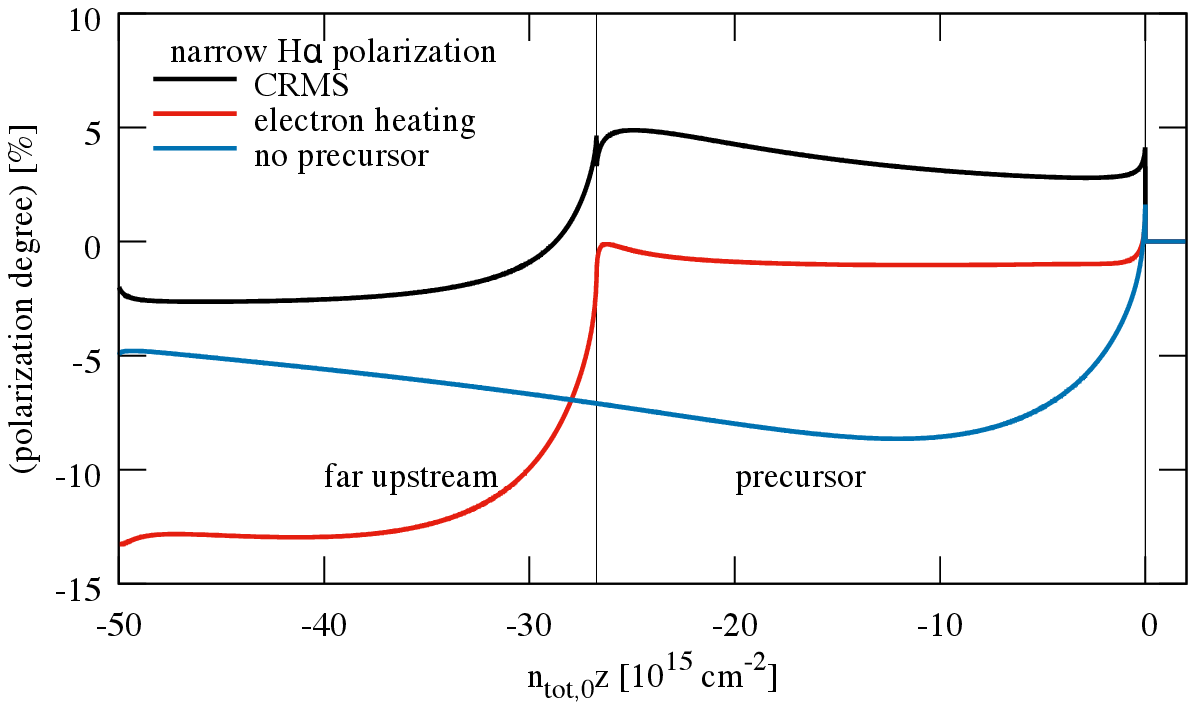}
\caption{
The polarization degree of H~$\alpha$.
The blue, red and black lines indicate the cases of no precursor (i),
only an electron heating precursor (ii), and the electron heating precursor with decelerated protons (iii),
respectively.
The vertical thin lines at $n_{\rm tot,0} z\simeq-27\times10^{15}~{\rm cm^{-2}}$
and $n_{\rm tot,0} z\simeq0~{\rm cm^{-2}}$ indicate the locations of the precursor front
and the shock front, respectively.
Positive values correspond to polarization along the $z$-axis,
while negative values correspond to polarization along the $x$-axis.
Note that these results includes all of atomic levels and processes
unlike the result shown in Figure~\ref{fig:simple}.}
\label{fig:polarization}
\end{figure}
%
Figure~\ref{fig:polarization} shows the estimated polarization degree of
H~$\alpha$ in the upstream region. Here the Stokes parameters are integrated
over an interval of frequency corresponding to the Doppler velocity range
$-25$ to $25~{\rm km~s^{-1}}$. As discussed in Section~\ref{sec:scattering},
the sign of Stokes $Q$ depends on the velocity modification. In the no
precursor case (i), the polarization degree is positive ($\simeq2$ per cent)
in the region close to the shock front where the optical depth of Ly~$\beta$
is small. Then, the polarization becomes negative at optical depth $\sim1$.
Since the Ly~$\beta$ radiated from the downstream region in the direction of
$\mu\approx-1$ (along the $z$-axis) is not so attenuated compared to that
radiated along $\mu\approx0$, the `photon beam' is elongated along the
$z$-axis, giving a negative polarization in the region distant from the
shock. Note that the polarization at $z\simeq-5\times10^{16}$~cm is affected
by the photon free escape boundary. On the other hand, in the simple electron
heating precursor case (ii), the polarization degree is modest ($\simeq-1$
per cent) in the precursor region. This is because the electron heating
precursor with no velocity modification yields a uniform, isotropically
emitting medium whose polarization property is shown in
Figure~\ref{fig:simple}. Note that the polarization degree in front of the
shock is $\simeq1$ per cent in this case (the plots in
Figure~\ref{fig:polarization} overlap). In the modified shock case (iii), the
polarization is positive with a degree of $\simeq5$ per cent. The degree of
attenuation of Ly~$\beta$ photons radiated in the direction $\mu\approx1$
depends on whether the decelerated/non-decelerated atoms emit/absorb them,
while the photons radiated in the direction $\mu\approx0$ are attenuated by
the both populations of atoms. Thus, the photons with $\mu\approx0$ are
efficiently attenuated resulting in positive polarization.
\par
%
\begin{figure}
\includegraphics[scale=0.7]{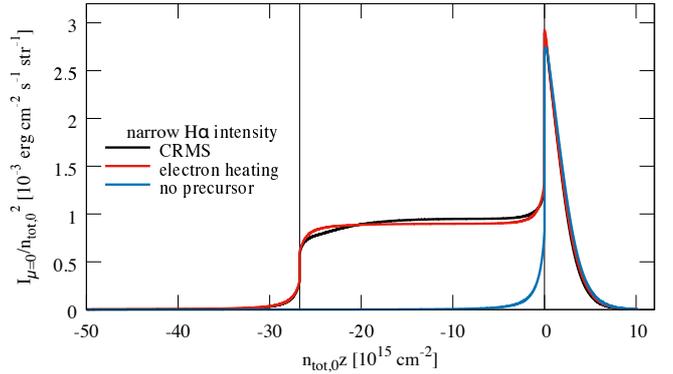}
\caption{
The surface brightness profile of H~$\alpha$.
The blue, red and black lines indicate the cases of no precursor (i),
only the electron heating precursor (ii), and the electron heating precursor with decelerated protons (iii),
respectively.
The vertical thin lines at $n_{\rm tot,0} z\simeq-27\times10^{15}~{\rm cm^{-2}}$
and $n_{\rm tot,0} z\simeq0~{\rm cm^{-2}}$ indicate the locations of the precursor front
and the shock front, respectively.}
\label{fig:brightness}
\end{figure}
%
Figure~\ref{fig:brightness} shows the surface brightness profiles of the
frequency-integrated Stokes $I$ for the three cases. In the existence of a
precursor, cases (ii) and (iii), the emission from the upstream region is
bright and comparable to that from the downstream region. Thus, in an actual
observation, the polarization of such precursor emission is detectable.
Indeed, \citet{sparks15} discovered the polarized H~$\alpha$ with a
polarization degree of $2.0\pm0.4$~per cent at SNR~SN~1006. Note that the
precursor emissions were reported in several cases in the
literature~\citep[e.g.][]{lee07, katsuda16,knezevic17} although their origin
is still a matter of debate. Candidates for the origin of
these precursor emissions often discussed are the CR precursor, a
photoionization precursor and a fast neutral precursor as the result of
charge exchange reactions~\citep[e.g.][]{morlino12}. In particular, by
constructing an H~$\alpha$ emission model with a fast neutral precursor,
\citet{morlino12} showed that the fast neutral precursor can contribute up to
$\sim40$ per cent of the total H~$\alpha$ emission in the case of complete
proton-electron temperature equilibration and $V_{\rm sh}\approx2500~{\rm
km~s^{-1}}$. Note that the fast neutral precursor also leads to an upstream
velocity modification of $\sim10$ per cent of $V_{\rm sh}$~\citep{blasi12,ohira12}
for $V_{\rm sh}\la2000~{\rm km~s^{-1}}$. For $V_{\rm sh}\ga2000~{\rm
km~s^{-1}}$, the modification hardly occurs because of the rapid decrease of
the charge-exchange cross-section at relative velocities $\ga2000~{\rm
km~~s^{-1}}$.
%
\begin{figure}
\includegraphics[scale=0.7]{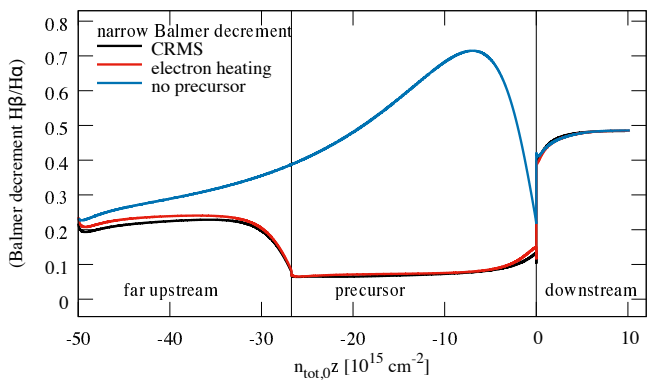}
\caption{
The Balmer decrement (total intensity ratio of H~$\beta$ to H~$\alpha$).
The blue, red and black lines indicate the cases of no precursor (i),
only the electron heating precursor (ii), and the electron heating precursor with decelerated protons (iii),
respectively.
The vertical thin lines at $n_{\rm tot,0} z\simeq-27\times10^{15}~{\rm cm^{-2}}$
and $n_{\rm tot,0} z\simeq0~{\rm cm^{-2}}$ indicate the locations of the precursor front
and the shock front, respectively.}
\label{fig:decrement}
\end{figure}
%
\par
Figure~\ref{fig:decrement} is the ratio of the frequency-integrated Stokes
$I$ of H~$\beta$ to that of H~$\alpha$. The ratio does not depend on the
existence of a velocity modification, in other words, the net optical depth
of the Lyman lines does not depend on the modification. This is also
indicated by Figure~\ref{fig:crpr}. The reason is that both decelerated and
non-decelerated hydrogen atoms survive sufficiently in the precursor region
for the absorption of Lyman lines. Thus, the polarimetry of H~$\alpha$ is a
unique diagnostic for the velocity modification of a shock that is one of the
essential predictions for collisionless shocks efficiently accelerating
non-thermal particles.
\par
The intensity ratio of H~$\beta$ to H~$\alpha$ (the
so-called Balmer decrement) is often discussed as a ratio in photon counts,
which is obtained by multiplying the intensity ratio by a factor of
0.74~\citep[see][]{shimoda19}. \citet{tseli12} evaluated the ratio in photon
counts as $\simeq0.3$ for the case of only proton direct collisional
excitation by assuming Delta-function distributions for protons and hydrogen
atoms. Note that this estimate does not include the Lyman-Balmer conversions
and cascades from higher levels. Following \citet{tseli12}, if we evaluate
the downstream ratio in photon counts including the electron direct
collisional excitation for a typical relative collision velocity of
$\simeq3/4V_{\rm sh}=3000~{\rm km~s^{-1}}$, we obtain $\simeq0.38$. Thus, the
ratio at the edge of downstream, $\simeq0.5\times0.74\simeq0.37$, may be
almost determined by the collisional excitation for typical relative
velocities. Toward the shock front, the ratio becomes smaller due to the
Lyman-Balmer conversions. Note that because the oscillator strength of the
transition from 1s to 3p is stronger than that to 4p, the
Ly~$\beta$-H~$\alpha$ conversion is more efficient than the
Ly~$\gamma$-H~$\beta$ conversion.

\section{Summary and Discussion}
\label{sec:discussion} We have shown that the polarimetry of H~$\alpha$ can
be a useful and unique diagnostic of a CR modified shock. The H~$\alpha$
emission from the upstream region mainly originates from the
Ly~$\beta$-H~$\alpha$ conversion as a consequence of photon-scattering, even
if there is an electron heating precursor yielding collisional excitation.
Therefore, the upstream H~$\alpha$ is linearly polarized with a polarization
degree of $\sim$ a few per cent. The polarization direction (angle) depends
strongly on whether there is a velocity modification. For the case of no
velocity modification, the polarization direction is parallel to the shock
surface. On the other hand, if there are velocity modified hydrogen atoms
generated by charge-exchange reactions with the decelerated upstream protons,
the direction is perpendicular to the shock surface. Moreover, the upstream
surface brightness of H~$\alpha$ is comparable to the downstream brightness.
Furthermore, we have shown that even if the velocity modification is just 5
per cent of $V_{\rm sh}$, the polarization of H~$\alpha$ responds to the
modification sensitively. Hence, polarimetry of upstream H~$\alpha$ may be
realised that will constrain the modification of the shock due to the CR back
reactions and bring new insights to particle acceleration in collisionless
shocks.
\par
There is another diagnostic for the CR acceleration rate that is often
discussed. H~$\alpha$ emission with full width at half maximum (FWHM) of
$30\mathchar`-50~{\rm km~s^{-1}}$ is often observed in
SNRs~\citep[e.g.][]{sollerman03} and it indicates too high a temperature of
$2.5\mathchar`-5.6~{\rm eV}$ for hydrogen atoms to exist if the case of
thermal equilibrium holds in the preshock medium. Thus, this anomalous width
implies that there is non-thermal pre-heating before shock passage. If the CR
acceleration efficiency is really high, such non-thermal heating can occur,
for example, via generation and damping of Alfv\`enic turbulence in the
upstream region. Note that the CR back reaction mainly affects charged
particles in the plasma but the neutral atoms can be coupled with the
affected ions by charge-exchange reactions. \citet{morlino13} demonstrated
this by a semi-analytical model and showed that the width depends on the
maximum energy of the accelerated CRs. On the other hand, \citet{ohira16}
reported no pre-heating in his hybrid simulation, although his hybrid
simulation was generally limited to too short a time scale compared to
an actual SNR. Note that his simulation showed the
velocity modification with $\sim10$ per cent of $V_{\rm sh}$. Moreover, for
the Cygnus Loop SNR, \citet{medina14} pointed out that the FWHM of
$\simeq30~{\rm km~s^{-1}}$ can be realized in the upstream region by the
photo-ionization heating due to radiation from the post-shock or SNR ejecta.
This would be accentuated by the photoionization of H$_2$ molecules, followed
by dissociative recombination, whereby the neutral atoms fly apart with the
correct energy to produce the observed line width.
H$_2$ emission is observed in the northeast part of Cygnus Loop
though it can be from a radiative shock~\citep{graham91a,graham91b}.
Furthermore, \citet{shimoda19} showed that
the interaction between the SNR shocks and a dense upstream medium
$\ga30~{\rm cm^{-3}}$ can also result in a $30\mathchar`-50~{\rm km~^{-1}}$
FWHM due to the scattering of H~$\alpha$. Thus, we should still consider
carefully whether the width of H~$\alpha$ is really firm evidence of
efficient CR acceleration.

\section*{Acknowledgements}

This work is partially supported by JSPS KAKENHI grant no. JP18H01245.
JML was supported by the Guest Investigator Grant HST-GO-13435.001 from the
Space Telescope Science Institute and by the NASA Astrophysics Theory Program
(80HQTR18T0065), as well by Basic Research Funds of the CNR.




\bibliographystyle{mnras}
\bibliography{mnras_sl19b}




\appendix





\bsp	
\label{lastpage}
\end{document}